\documentclass[%
 aip,
 amsmath,amssymb,
preprint,
]{revtex4-1}
\usepackage{xcolor}
\usepackage{graphicx}
\usepackage{dcolumn}
\usepackage{bm}

\usepackage[utf8]{inputenc}
\usepackage[T1]{fontenc}
\usepackage{mathptmx}
\usepackage{etoolbox}

\makeatletter
\def\@email#1#2{%
 \endgroup
 \patchcmd{\titleblock@produce}
  {\frontmatter@RRAPformat}
  {\frontmatter@RRAPformat{\produce@RRAP{*#1\href{mailto:#2}{#2}}}\frontmatter@RRAPformat}
  {}{}
}%
\makeatother
\begin{document}

\preprint{AIP/123-QED}

\title{Hybridized magnonic materials for THz frequency applications.}
\author{D.-Q. To}%
\affiliation{Department of Materials Science and Engineering, University of Delaware, Newark, Delaware 19716, USA}%

\author{A. Rai}
\affiliation{Department of Physics and Astronomy, University of Delaware, Newark, Delaware 19716, USA}%

\author{J. M. O. Zide}%
\affiliation{Department of Materials Science and Engineering, University of Delaware, Newark, Delaware 19716, USA}%

\author{S. Law}%
\affiliation{Department of Materials Science and Engineering, University of Delaware, Newark, Delaware 19716, USA}%
\affiliation{Department of Materials Science and Engineering, The Pennsylvania State University, University Park, Pennsylvania 16802, USA}%

\author{J. Q. Xiao}%
\affiliation{Department of Physics and Astronomy, University of Delaware, Newark, Delaware 19716, USA}%

\author{M. B. Jungfleisch}%
\affiliation{Department of Physics and Astronomy, University of Delaware, Newark, Delaware 19716, USA}%
\address{Authors to whom correspondence should be addressed: mbj@udel.edu, doty@udel.edu}

\author{M. F. Doty}%
\affiliation{Department of Materials Science and Engineering, University of Delaware, Newark, Delaware 19716, USA}
\address{Authors to whom correspondence should be addressed: mbj@udel.edu, doty@udel.edu}

\date{\today}

\begin{abstract}
The capability of magnons to hybridize and strongly couple with diverse excitations offers a promising avenue for realizing and controlling emergent properties that hold significant potential for applications in devices, circuits, and information processing. In this letter, we present recent theoretical and experimental developments in magnon-based hybrid systems, focusing on the combination of magnon excitation in an antiferromagnet with other excitations, namely plasmons in a topological insulator, phonons in a 2D AFM, and photons. The existence of THz frequency magnons, plasmons, and phonons makes magnon-based hybrid systems particularly appealing for high-operating-speed devices. In this context, we explore several directions to advance magnon hybrid systems, including strong coupling between a surface plasmon and magnon polariton in a TI/AFM bilayer, a giant spin Nernst effect induced by magnon phonon coupling in 2D AFMs, and control of magnon-photon coupling using spin torque.
\end{abstract}

\maketitle

Magnons are elementary, collective, and charge-neutral excitations of localized spins (i.e.~spin waves) that exhibit bosonic quasiparticle behavior. Magnons obey the Bose-Einstein distribution function at finite temperatures and possess a zero chemical potential in equilibrium. The absence of charge flow and the existence of THz frequency excitations makes magnons an intriguing potential medium for high-speed information transmission devoid of energy dissipation\cite{Neusser2009,Barman2021,Chumak2015}. However, despite significant research efforts, the controlled generation of THz magnons in magnetic materials remains a challenge, typically relying on thermal source conversion \cite{Han2019,Fulara2019,Wang2019}. This challenge is in many ways a materials problem. Ferromagnetic materials (FMs) exhibit a net magnetization that makes them useful for various information storage applications such as magnetic random-access memories (MRAMs) and computer hard disks \cite{tsymbal2019spintronics,hirohata2020review}. FMs are also used in emerging fields such as artificial intelligence, neuromorphic computing, and in-memory computing, which are currently witnessing an explosive growth \cite{hirohata2020review,grollier2020neuromorphic}. In contrast, antiferromagnetic materials (AFMs) possess zero net magnetization due to their antiparallel-aligned magnetic sublattices. AFMs have several advantages over FMs for high-speed spintronic and information transmission applications, including (1) three orders of magnitude enhanced operation speed due to stronger exchange coupling between neighboring spins (of order meV for AFMs in contrast to of order 10$^{-3}$~meV for FMs), (2) immunity to electromagnetic interference, and (3) a much-enhanced memory density \cite{jungwirth2016,baltz2018,vzelezny2018,siddiqui2020}. However, the challenges to realizing AFM-based spintronic devices are daunting, primarily because AFMs nearly completely lack both physical observables (resistance, voltage, etc.) that are related to their magnetic order parameter, known as the Néel vector, and  the means to manipulate it \cite{jungwirth2016,baltz2018,vzelezny2018,siddiqui2020}. Understanding the dynamics of magnons in AFMs is crucial for determining the Néel vector switching characteristics and exploring electromagnetic applications at THz frequencies. 

A range of device opportunities have been proposed to exploit the unique properties of AFMs. For example, the characteristic THz resonant modes of AMFs have been predicted to enable the development of narrowband THz spintronic emitters driven by spin-orbit torques \cite{ChengRan2016}. Recent work has also demonstrated that off-resonant optical torque can excite spin current in AFMs, resulting in broadband THz emission in AFM thin films \cite{Qiu2021}. The development of ultrafast, nanoscale magnonic logic circuits also calls for new tools and materials to generate coherent magnons with high frequencies and short wavelength \cite{Chumak2015}, further emphasizing the importance of antiferromagnetic platforms.  While these areas of research are rapidly evolving, a less-explored domain involves the emergence of hybridized states. In this letter, we introduce a multifaceted exploration into the advancement of magnon-based hybrid systems, which emerge due to strong coupling between magnons and other elementary excitations, including plasmons, photons, and phonons in a heterostructure \cite{To2022,To2023a,To2023b,Rai2023}. The hybridized states are composed of the magnetic, photonic, plasmonic, or phononic resonances of the constituent materials and present an effective means to excite, manipulate, and detect magnons through coherent control of the hybridized states. This approach holds promise for the emergence of novel properties and the expansion of the spectrum of device applications and opportunities in this exciting and rapidly-evolving field \cite{Lachance2019,Li2020,Bhoi2020,Xu2021,Jostein2023,Anna2023}. For example, hybridized states may create avenues for magnetically tunning plasmons or optically controlling the Néel vector in AFMs. In this Letter we combine summaries of prior work with new computational results to provide a broad picture of the physical origins and consequence of magnon coupling in a range of hybrid materials. We first illustrate the intricate interactions between surface plasmon polaritons in both 2D and 3D topological insulators and magnon polaritons in antiferromagnetic (AFM) materials \cite{To2022,To2023a}. We next present new results that explore the manipulation and regulation of magnon-photon coupling through the ingenious utilization of spin torque mechanisms \cite{Rai2023}. Finally, we review the emergence of the colossal spin Nernst effect arising from the dynamic interaction between magnons and phonon quasiparticles in 2D AFMs \cite{To2023b}.

\begin{figure}[h]
\centering
    \includegraphics[width= 1\textwidth]{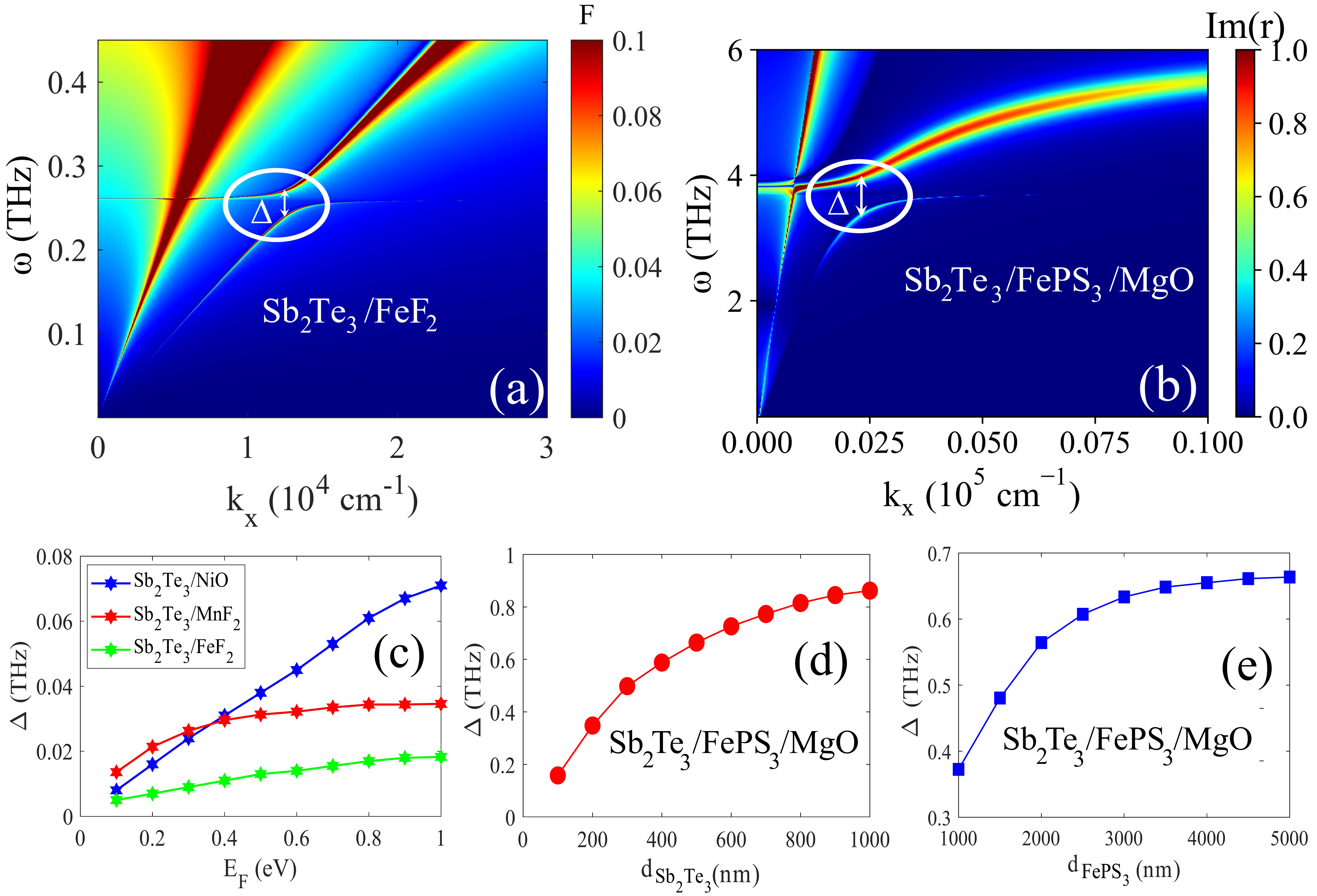}
 \caption{The dispersion relation of the surface Dirac plasmon-phonon-magnon polariton in the (a) Sb$_2$Te$_3$/FeF$_2$ structure and (b) Sb$_2$Te$_3$/FePS$_3$/MgO structure with $\Delta$ being the coupling strength parametrized by the gap at the resonant point. The coupling strength as a function of (c) the Fermi energy of surface plasmon in the TIs; (d) the TI's thickness and (d) the AFM's thicness.}
  \label{FIG1}
\end{figure}

Surface Dirac plasmon polaritons (DPP) are electromagnetic collective modes of electrons, characterized by their localization as evanescent waves perpendicular to the surface and propagating along the surface of a topological insulator (TI) \cite{Ginley2018a,Ginley2018b,Wang2020,To2022b}. When one combines materials like TIs and magnetic substances to create hybrid structures, an incident electromagnetic (EM) wave can excite the internal degrees of freedom within each constituent material. This leads to the emergence of collective excitations, known as polaritons, that may provide novel optical and electrical functionalities \cite{Zhang2012,Torma2014,Bludov2019,Dantas2023,Wang2023}. For instance, Dirac surface-plasmon phonon magnon polaritons (DPPMPs) may emerge in a TI/AFM bilayer when the interaction between surface plasmons in the TI and magnons in the AFM surpasses a critical threshold, thereby pushing the system into a regime known as \textit{strong coupling} \cite{To2023a,Costa2023}. The distinctive hallmark of this hybridized state is typically observed in the dispersion relation as an anti-crossing, also referred to as an avoided crossing. Such anticrossings are indicated by the circles in Fig.~\ref{FIG1}(a) and (b). The strength of the interaction can be quantified by examining the amplitude of the avoided-crossing splitting between the two polariton branches, indicated by $\Delta$ in Fig.~\ref{FIG1}(a) and (b). By analogy to the field of cavity quantum electrodynamics, the regime of strong coupling can be described by the regime in which the cooperativity factor $C=\frac{\Delta^{2}}{4\Gamma_{1} \Gamma_{2}} \geq 1$. In this equation, $\Delta$ represents the energy gap between the coupled polariton states, while $\Gamma_{1}$ and $\Gamma_{2}$ denote the line widths associated with the isolated excitations comprising the hybridized states. These line widths arise from the inherent loss or dissipation associated with each excitation. 

To obtain the dispersion of these surface-plasmon magnon polaritons in a TI/AFM heterostructure, semiclassical techniques based on Maxwell's equations can be employed. Maxwell's equations can be effectively solved using well-established methods such as the transfer matrix or scattering matrix approach \cite{To2022b,To2022}. Please refer to supplemental material (SM) for the details of the method used here. Employing this semiclassical approach allows for the comprehensive characterization of the properties of surface DPPMPs as a function of the structural and material attributes of the constituent materials. This, in turn, enables the quantification of the coupling strength between the topological insulator and AFM material through the magnetic degree of freedom. This methodology provides us with a deeper understanding of the underlying physics behind the coupling constant and empowers us to predict potential material combinations that could exhibit even stronger coupling effects.

Figure \ref{FIG1}(a) illustrates the dispersion of DPPMPs within the Sb$_{2}$Te$_3$/FeF$_2$ bilayer, with FeF$_2$ serving as a semi-infinite layer. The presence of DPPMPs in this configuration is denoted by the appearance of an avoided crossing point at the magnonic resonance in FeF$_2$ occurring at wavevector $k_x = 1.4 \times 10^{4}\text{cm}^{-1}$ and frequency $\omega = 0.25\text{THz}$ (indicated by the white circle).  The strength of this coupling, characterized by the gap $\Delta$ shown in the figure, can be adjusted by varying the Fermi energy (E$_F$) of surface plasmons in the topological insulator (TI), as depicted in Figure \ref{FIG1}(c). This dependence of $\Delta$ on E$_F$ arises because the plasmon polariton frequency is a function of the Fermi level, and thus changing E$_F$ can shift the surface plasmon polariton in the TI towards resonance with the magnon polariton in the antiferromagnetic (AFM) material, increasing the strength of the coupling [see section I(A) of the SM for detail]. Figure \ref{FIG1}(c) also illustrates the enhanced coupling strength when switching from NiO to FeF$_2$ as the AFM material. This underscores the fundamental role of the AFM material properties in governing the interaction between the TI and AFM. In this context, it is the magnetic dipole moment in the AFM material, which is directly proportional to the anisotropy energy, that plays a crucial role. Therefore, pursuing an AFM material with a higher anisotropy energy promises a significant enhancement in coupling strength \cite{To2022,To2023a}.

Figure \ref{FIG1}(b) presents the DPPMPs in the Sb$_{2}$Te$_3$/FePS$_3$ system, where the anisotropy energy in FePS$_3$ is approximately three orders of magnitude greater than that in FeF$_2$. Consequently, the coupling strength in this structure is predicted to be roughly one order of magnitude larger, potentially reaching the strong coupling regime in which the cooperativity factor exceeds unity. The larger coupling strength occurs because the anisotropy energy of an antiferromagnetic (AFM) material determines its magnetic dipole [see section I(B) of the SM for detail]. A larger magnetic dipole, in turn, leads to a stronger interaction between the electromagnetic wave's magnetic component and the local spin moment in the AFM. This increased interaction results in the excitation of magnon polaritons with more magnons. Consequently, there is a larger contribution of magnons to the formation of Dirac plasmon phonon magnon hybrid modes, strengthening the interaction between magnon states in the AFM and Dirac plasmon phonon states in the topological insulator (TI).  This analysis suggests that any AFM material with anisotropy energy comparable to that of FePS$_3$ (of order one meV) may be a promising alternative candidate for realizing strong coupling between a surface-plasmon-phonon polariton in a TI and magnon polaritons in an AFM.

The dependence of the coupling strength on various structural parameters is revealed in Figure 1(d) and (e), showcasing a growing coupling strength as the thickness of the topological insulator (TI, panel d) and antiferromagnetic (AFM, panel e) layers increase. These phenomena can be primarily attributed to two key factors. First, the contribution of magnons to the hybridized state increases as the AFM layer thickness is increased because more magnon modes actively participate in the hybridized mode \cite{To2023a}. Second, the surface plasmon (SP) mode shifts as the TI thickness increases, increasing the spectral overlap of the surface plasmon and magnon modes. This shift occurs because the surface plasmon modes in the TI are coupled modes containing contributions from the plasmons on the top and bottom surfaces, resulting in a thickness-dependent surface plasmon mode energy. We note that the relationship between surface plasmons and TI thickness is mediated by the phonon in the TI, which causes the dielectric function of the TI to become negative at high frequencies. Therefore, selecting a TI material with a robust interaction between plasmons and phonons provides another important tool.

In summary, the pursuit of strong coupling between surface plasmons and magnon polaritons typically hinges on selecting a TI with a large plasmon-phonon interaction and an AFM material with substantial anisotropy energy. One promising candidate for achieving this synergy is the Sb$_{2}$Te$_3$/FePS$_3$ structure, as demonstrated here. Possible alternative AFM material that are promising include: L1$_2$ IrMn$_3$, Na$_4$IrO$_4$, and Cr–trihalide Janus monolayers with applied strain up to 5$\%$ \cite{To2023a}.


Ferromagnetic magnons and superconducting qubits have resonances at microwave frequencies enabling the creation of magnonic hybrid systems that promise to advance quantum information technologies due to the flexibility and tunability of magnons \cite{ArtmanPRB1953,SoykalPRL2010,HueblPRL2013,ZhangPRL2014,TabuchiPRL2014,TabuchiScience2015,BaiPRL2015,ChumakIEEE2022}. Most studies on coherent magnon-photon coupling physics used yttrium iron garnet (YIG) as a source of magnons due to its very low Gilbert damping \cite{ChangIEEE2014,DingIEEE2020} and high spin density \cite{PhysRevGilleo1958}, which are necessary to enter the strong coupling regime. To enhance the coupling between magnons and photons, different designs of three-dimensional cavities \cite{ArtmanPRB1953,ZhangPRL2014,HarderPRL2018}, split-ring resonators \cite{BhoiSReports2017,BhoiPRB2019} and superconducting resonators \cite{LiPRL2019} have been used. Having an external parameter to efficiently control magnon-photon coupling is essential, which has motivated exploration of other ferromagnetic materials as a magnon source. For example, metals offer easier miniaturization than YIG and magnons in metals can be induced by, for example, spin torque. Having an external control parameter such as spin torque could also enable toggling between coherent (level repulsion) and dissipative coupling (level attraction). One of the distinct features of dissipative coupling is the presence of exceptional points where the eigenstates and eigenvalues of the hybrized modes coalesce. Systems at exceptional points tend to respond more drastically to external perturbations, which makes this effect interesting for developing ultrasensitive sensors \cite {WiersigPRL2014,WiersigPRA2016,ChenNanture2017,HodaiNature2017}

Extending the model introduced by Rai et al. \cite{Rai2023}, we can find spin-torque control of the dissipative coupling described by the following equation,

\begin{equation}
 \tilde{\omega}_{\pm}=\frac{\left(\frac{\omega_{c}}{1+i \beta}+\frac{\omega_{m}-\delta}{1+i \alpha}\right) \pm \sqrt{\left(\frac{\omega_{c}}{1+i \beta}-\frac{\omega_{m}-\delta}{1+i \alpha}\right)^{2}-\frac{2 \omega_{c} \omega_{s} K_{F} K_{L}}{(1+i\alpha)(1+i\beta)}}}{2},
 \label{MP coupling}
\end{equation}
where $\delta=\gamma\tilde{c}_{J}$. Here, $\gamma$ is the gyromagnetic ratio and $\tilde{c}_{J}$ is a complex term associated with the damping-like torque coefficient ($b_\mathrm{J}$) and field-like torque coefficient ($a_\mathrm{J}$) defined by $\tilde{c}_\mathrm{J}=b_\mathrm{J}-ia_\mathrm{J}$. $\omega_{c}$ is the cavity frequency, $\omega_{m}$ is the magnon frequency, and $\omega_{s} = \gamma M_{s}$ ($M_{s}$ is the saturation magnetization). $\alpha$ and $\beta$ are the dissipation rates associated with the magnon and photon modes, respectively. $K_{F}$ and $K_{L}$ are the coupling terms associated with the Faraday and Lenz effects, respectively \cite{HarderPRL2018}. Here, $\tilde{\omega}$ can be expanded in the form $\tilde{\omega}=\omega - i\Delta\omega$, where $\omega$ is the eigenfrequency and $\Delta\omega$ is the linewidth. Equation~(\ref{MP coupling}) can be obtained by coupling the generalized Landau-Lifshitz-Gilbert equation with the RLC equations describing the cavity \cite{HarderPRL2018,Rai2023}.

\begin{figure}[h]
\centering
\includegraphics[width = 0.9\textwidth, keepaspectratio]{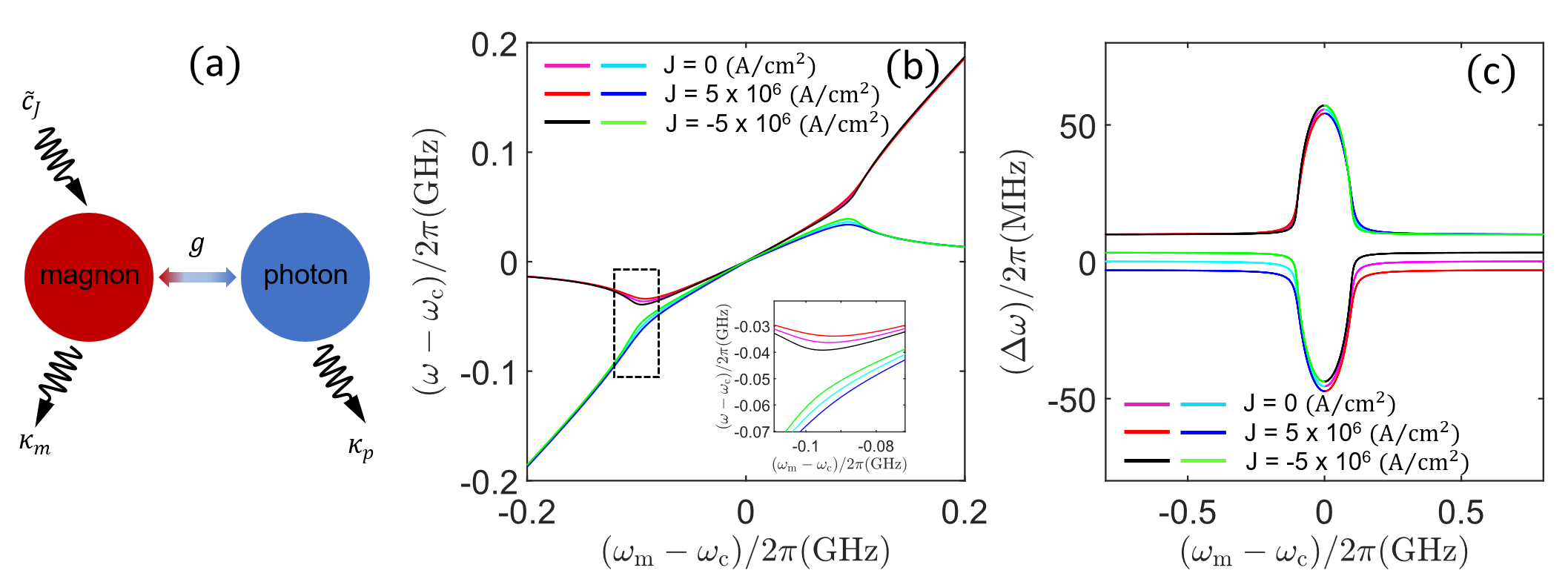}
\caption{(a) Schematic illustration of spin-torque manipulation of magnon-photon coupling. $g$, $\kappa_{m}$, and $\kappa_{p}$ represent the coupling strength between the magnon and photon modes, dissipation rate of magnon modes, and the dissipation rate of photon modes, respectively. $\tilde{c}_\mathrm{J}\left( = b_{J}-ia_{J}\right)$ is the term associated with field-like ($b_{J}$) and damping-like torques ($a_{J}$). The dispersion ($\omega-\omega_{c}$) in (b) and the linewidth ($\Delta{\omega}$) in (c) are plotted as a function of the field detuning ($\omega_{m}-\omega_{c}$) for $\alpha = 1\times10^{-3}$. We compare the hybridized magnon-photon mode for different dc current densities $J$: The magenta and cyan lines, red and blue lines, and black and green lines represent the two hybridized modes for $J=0~\mathrm{A/cm^2}$, $J=-5\times10^6~\mathrm{A/cm^2}$, and $J=5\times10^5~\mathrm{A/cm^2}$, respectively.  The inset in (b) is the zoomed in part of the marked rectangle.}
\label{fig:spintorque}
\end{figure}

Equation~(\ref{MP coupling}) \cite{Rai2023} explains the effect of spin torque on the coherent magnon-photon coupling. This model can be extended to the dissipative magnon-photon coupling regime by replacing the coupling coefficient $K_{F}K_{A}$ by -$K_{F}K_{L}$, leading to a level attraction \cite{HarderPRL2018}. This means that the combined effect of the Faraday and Lenz effects ($K_{F}K_{L}$) opposes the combined effect of the Faraday and Amperes effects ($K_{F}K_{A}$) \cite{HarderPRL2018}. The term $K_{F}K_{A}$ results in a coherent magnon-photon coupling, whereas the term -$K_{F}K_{L}$ gives the dissipative magnon-photon coupling. Here, the negative sign is due to the opposing role of the Lenz effect. Figure~\ref{fig:spintorque}(a) schematically illustrates the effect of spin torques on the magnon-photon coupling process. The following results are obtained using a cavity frequency $\omega_{c}/2\pi=10~\mathrm{GHz}$ with a cavity dissipation rate $\beta=1\times10^{-5}$. The reduced gyromagnetic ratio ($\gamma/2\pi$), damping-like torque efficiency ($\eta_{a}$), and field-like torque efficiency ($\eta_{b}$)  are taken as $2.8\times10^{6}~\mathrm{Hz/Oe}$, $0.2$, and $0.05$, respectively. As a magnetic material, we choose YIG with a film thickness $t=2\times10^{-5}~\mathrm{cm}$ and saturation magnetization, $M_{s}=144~\mathrm{emu/cm^3}$,~ \cite{MojtabaQST2022}. For the calculation, the term $K_{F}K_{L}$ is taken as $-10\times10^{-5}$ and $\alpha$ is taken as $1\times10^{-3}$,~ \cite{PirroAPL2013,BenjmainJAP2014,BenjaminPRB2015,HaertingerPRB2015}.

For the analysis, we choose three values of current density:  $J=-5\times10^6~\mathrm{A/cm^2}$, $J=0~\mathrm{A/cm^2}$, and  $J=5\times10^6~\mathrm{A/cm^2}$. As can be seen from Fig.~\ref{fig:spintorque}(b), the transition into the phase where two hybridized modes coalesce for all three values of $J$ is not abrupt, as one would expect for exceptional points. The main reason behind this is the relatively large value of $\alpha$ taken into consideration for the analysis. We note that the two hybridized modes come closest for $J=-5\times10^6~\mathrm{A/cm^2}$ and separate farther away for $J=5\times10^6~\mathrm{A/cm^2}$ (see the inset of Fig.~\ref{fig:spintorque}(b)), indicating that the magnitude and direction of the dc current density can be used to enhance the effect of dissipative coupling. The linewidth plot shown in Fig.~\ref{fig:spintorque}(c) clearly indicates the presence of the dissipative coupling (repulsion of the linewidths) for all values of $J$. The important implication of these results is that efficient control of the dissipative coupling process is possible with current densities of the order $10^6$~A/cm$^2$ and a moderate Gilbert damping of the order $10^{-3}$. This suggests a path toward more robust dissipative coupling, opening new avenues for quantum sensors.

\begin{figure}[h]
\centering
    \includegraphics[width= 0.9\textwidth]{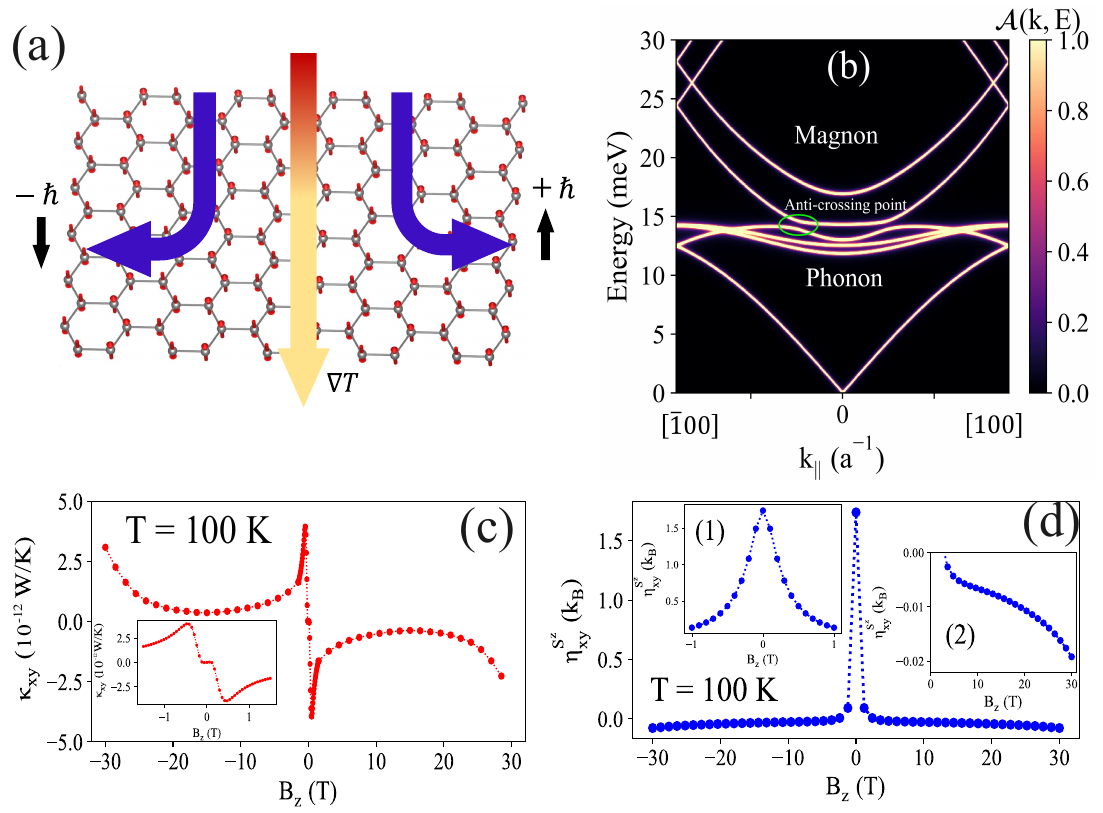}
 \caption{(a) Schematic view of the magnon SNE in a 2D AFM where  transverse flow of magnons carrying opposite out-of-the plane spins ($\pm ~\hbar$) is induced by a temperature gradient $\nabla T$ along the longitudinal direction. (b) Plot of the spectral function $\mathcal{A}\left(k,E \right)$ (color code) versus energy and wave vector along the [100]-direction showing the dispersion of the magnon-polaron system under an applied magnetic field of $B=30~$T in FePS$_3$. The anticrossing between the magnon- and phonon-like bands is indicated by the green circle; the existence of this anticrossing indicates strong coupling between the magnon and phonon excitations in this material. (c) The spin Nernst conductivities as a function of an applied magnetic field $B_z$ calculated at \mbox{$T=100$ K}. Two insets in panel (c) show a zoom in 
 for: (1) $B_{z} \in [-1~\mathrm{T},1~\mathrm{T}]$; and (2) $B_{z} \in [2~\mathrm{T},30~\mathrm{T}]$}
  \label{FIG2}
\end{figure}

The robust interplay between distinct excitations carries profound implications that can give rise to the emergence of unique and intriguing properties \cite{Park2020,Zhang2020,Bazazzadeh2021,Zhang2022}. For example, when we examine the dynamic interplay between magnons and phonons within a 2D antiferromagnetic (AFM) material such as FePS$_3$, we theoretically predict the manifestation of the thermal Hall effect (THE) and spin Nernst effect (SNE) mediated by magnon polarons \cite{To2023b}. These effects are not only observable, but also substantial in magnitude. Let us briefly summarize the conceptual foundation of the magnon THE and magnon SNE. The magnon THE refers to a phenomenon that occurs when a temperature gradient applied to a magnetic material generates transverse thermal transport of magnons, perpendicular to both the temperature gradient and magnetization. The magnon THE  results in a net magnon current, i.e. the accumulation of magnon at one side of the 2D material. The magnon SNE is analogous to the electronic Spin Hall Effect (SHE) in which electrons with opposite spins move in opposite directions perpendicular to the applied charge current. The magnon SNE involves the motion of magnons instead of electrons, with each magnon carrying spin polarization in opposite directions orthogonal to the temperature gradient, as depicted in Figure~\ref{FIG2}(a). The magnon SNE phenomenon arises due to the presence of two distinct magnon species in antiferromagnetic materials, each carrying an opposite spin polarization. Recent research has demonstrated that the magnon SNE effect can be observed in several scenarios: (1) Collinear antiferromagnets on a honeycomb lattice: In systems characterized by a honeycomb lattice, such as collinear antiferromagnets, the Dzyaloshinskii-Moriya interaction (DMI) acts on magnons in a manner analogous to how spin-orbit coupling (SOC) influences electrons in SHE. (2) Noncollinear antiferromagnets: Even in the absence of any SOC responsible for DMI, noncollinear antiferromagnets exhibit the magnon SNE effect, even in the absence of an applied magnetic field. (3) Collinear antiferromagnets or ferrimagnets with magnetoelastic coupling: In collinear antiferromagnets or ferrimagnetic materials, magnetoelastic coupling hybridizes the magnon and phonon quasiparticle bands. The regions of anticrossing in these bands are believed to be crucial for generating nonzero Berry and spin Berry curvature, which drive transverse transport in THE and SNE, respectively \cite{Zhang2022}.

In the case of 2D AFM FePS$_3$ characterized by a zigzag order, the magnon thermal Hall and magnon Spin Nernst Effect may occur due to the robust coupling between magnons and phonons, resulting in the formation of a magnon-polaron system \cite{Liu2021,Vaclavkova2021,Pawbake2022,Cui2023}. When subjected to an external magnetic field, magnons with opposing spins experience a Zeeman-like spin-splitting phenomenon, wherein one of the magnon bands shifts to overlap with the phonon band, as illustrated in Figure \ref{FIG2} (b) around 15~meV in the energy scale. The magnetoelastic coupling facilitates the interaction between magnon and phonon, leading to the emergence of anticrossing points between these two bands, as highlighted by the green circle in Figure \ref{FIG2} (b). These anticrossing points signify the hybridization of bands with magnon and phonon character. In turn, this hybridization between magnons and phonons within FePS$_3$ gives rise to a finite thermal Hall [Fig.~\ref{FIG2} (c)]  and spin Nernst effect [Fig.~\ref{FIG2} (d)] carried by magnon-polarons \cite{To2023b}. Notably, even without an externally applied magnetic field, a substantial spin Nernst conductivity (SNC) is predicted, owing to phonon-mediated interband transitions among magnons. Fig~\ref{FIG2} (c) and (d) illustrate the thermal Hall and spin Nernst conductivity as a function of the applied magnetic field (B), showing a remarkably high value of SNC at zero magnetic field. This SNC then rapidly diminishes when the applied magnetic field exceeds a critical value (\mbox{$B_z \gtrsim 2$ T}). This decrease is primarily due to the widening energy gap between the two magnon-like bands, which, in turn, suppresses interband transitions between them. 

The magnon SNE induced by magnon-phonon coupling that is discussed here manifests across a wide spectrum of material systems. For instance, it is also predicted to occur in transition-metal trichalcogenides exhibiting both Néel magnetic order, such as MnPS$_3$ and VPS$_3$, and zigzag order, as exemplified by CrSiTe$_3$, NiPS$_3$, and NiPSe$_3$ \cite{Bazazzadeh2021}. This emphasizes the robustness of magnon-phonon hybridization for the generation of magnon spin currents, which has significant implications for the development of spintronic devices.

Hybridized magnonic systems offer many potential advantages for THz frequency device applications, potentially allowing us to exploit and control THz frequency magnons, plasmons, phonons, and photons in 3D and 2D AFMs, topological insulators, and superconductors. We explored several promising emergent phenomena in hybrid materials, including the investigation of strong coupling between surface plasmons and magnon polaritons in a topological insulator/antiferromagnet bilayer, the control of magnon-photon coupling using spin torque, and the giant spin Nernst effect induced by magnon-phonon coupling in a 2D AFM. These examples highlight 1) the potential of magnon-based hybrid systems for advancing device and information processing technologies and 2) the importance of both understanding and controlling material properties and interactions in order to realize such technologies.

See the supplementary material for extensive discussion of the theoretical computation method employed for generating the data presented in Fig.~\ref{FIG1} and Fig.~\ref{FIG2}.

The authors would like to thank Branislav K. Nikoli\'c for helpful discussions. This research was primarily supported by NSF through the University of Delaware Materials Research Science and Engineering Center, DMR-2011824. Work on magnon-photon polariton control by spin torque was supported by the U.S. Department of Energy, Office of Basic Energy Sciences, Division of Materials Sciences and Engineering under Award DE-SC0020308.

\bibliography{biblio}

\end{document}